\pdfoutput=1
\documentclass[journal]{IEEEtran}
\usepackage{cite}

\ifCLASSINFOpdf
  \usepackage[pdftex]{graphicx}
\else
\fi
\usepackage{amsmath}
\usepackage{amssymb}
\usepackage{chemarrow}
\usepackage{array}

\ifCLASSOPTIONcompsoc
 \usepackage[caption=false,font=normalsize,labelfont=sf,textfont=sf]{subfig}
\else
 \usepackage[caption=false,font=footnotesize]{subfig}
\fi
\usepackage{float}
\usepackage{url}
\usepackage[colorlinks,linkcolor=black,anchorcolor=black,citecolor=black]{hyperref}

\usepackage{booktabs}
\usepackage{tabularx}
\usepackage{multirow}
\usepackage{diagbox}

\usepackage{calc}

\begin{document}
%
\title{Fine Perceptive GANs for Brain MR Image Super-Resolution in Wavelet Domain}
%
%
%
\author{Senrong You\textsuperscript{\rm 1, 2}, 
		Yong Liu\textsuperscript{\rm 3},
		Baiying Lei\textsuperscript{\rm 4},
		Shuqiang Wang\textsuperscript{\rm 1}%
\thanks{Corresponding author: Shuqiang Wang, Email:sq.wang@siat.ac.cn}
\thanks{\textsuperscript{\rm 1} Shenzhen Institutes of Advanced Technology, Chinese Academy of Sciences}%
\thanks{\textsuperscript{\rm 2} University of Chinese Academy of Sciences}%
\thanks{\textsuperscript{\rm 3} Renmin University of China}%
\thanks{\textsuperscript{\rm 4} Shenzhen University}}
\markboth{Manuscript}%
{Shell \MakeLowercase{\textit{et al.}}: Bare Demo of IEEEtran.cls for IEEE Journals}

%



\maketitle

\begin{abstract}

Magnetic resonance imaging plays an important role in computer-aided diagnosis 
and brain exploration. However, limited by hardware, scanning time and cost, 
it's challenging to acquire high-resolution (HR) magnetic resonance (MR) image clinically. 
In this paper, fine perceptive generative adversarial networks (FP-GANs) is proposed to 
produce HR MR images from low-resolution counterparts. It can cope with the detail insensitive 
problem of the existing super-resolution model in a divide-and-conquer manner. 
Specifically, FP-GANs firstly divides an MR image into low-frequency global approximation and 
high-frequency anatomical texture in wavelet domain. Then each sub-band generative adversarial 
network (sub-band GAN) conquers the super-resolution procedure of each single sub-band image. 
Meanwhile, sub-band attention is deployed to tune focus between global and texture information. 
It can focus on sub-band images instead of feature maps to further enhance the anatomical 
reconstruction ability of FP-GANs. 
In addition, inverse discrete wavelet transformation (IDWT) is integrated into model 
for taking the reconstruction of whole image into account. 
Experiments on MultiRes\_7T dataset demonstrate that FP-GANs outperforms the competing methods 
quantitatively and qualitatively.
\end{abstract}


%
\IEEEpeerreviewmaketitle

\section{Introduction}
%
%
%
%

\IEEEPARstart{M}{agnetic} resonance imaging (MRI) is a universally used medical imaging technology for 
auxiliary diagnosis of brain disease and brain function exploration. 
Compared with Computed Tomography (CT) and Positron Emission Tomography (PET), MRI provides 
clearer histopathological detail of soft tissue without cancer causing radiation exposure. 
However, due to physical constraints, MRI takes much longer time to acquire 
high-resolution (HR) magnetic resonance (MR) image clinically. It may disturb patients and 
inevitably lead to motion blur in MR image. 
There are generally two directions to cut down scanning time: 
\textbf{1)} Strengthen magnetic field with more advanced equipment. 
Magnetic field strength has been improved from low-field ($0.2$-$0.5$ Tesla) to 
ultra-high-field ($\geq 7$ Tesla) for higher signal noise ratio in many 
researches\cite{37_regatte2007ultra, 48_erturk2017toward, 38_geethanath2019accessible}. 
However, ultra-high-field MRI requires expensive hardware equipment and bring 
potential safety risk, which make it difficult to promote. 
\textbf{2)} Apply super-resolution (SR) algorithm. SR algorithms have achieved great success to 
increase the spatial resolution of MR images\cite{1_van2012super}. 
Moreover, it has been validated that computational approaches are more cost-effective and 
efficient for increasing spatial resolution\cite{39_plenge2012super}. 

\begin{figure}[htb]
	\centering
	\includegraphics[width=0.48\textwidth]{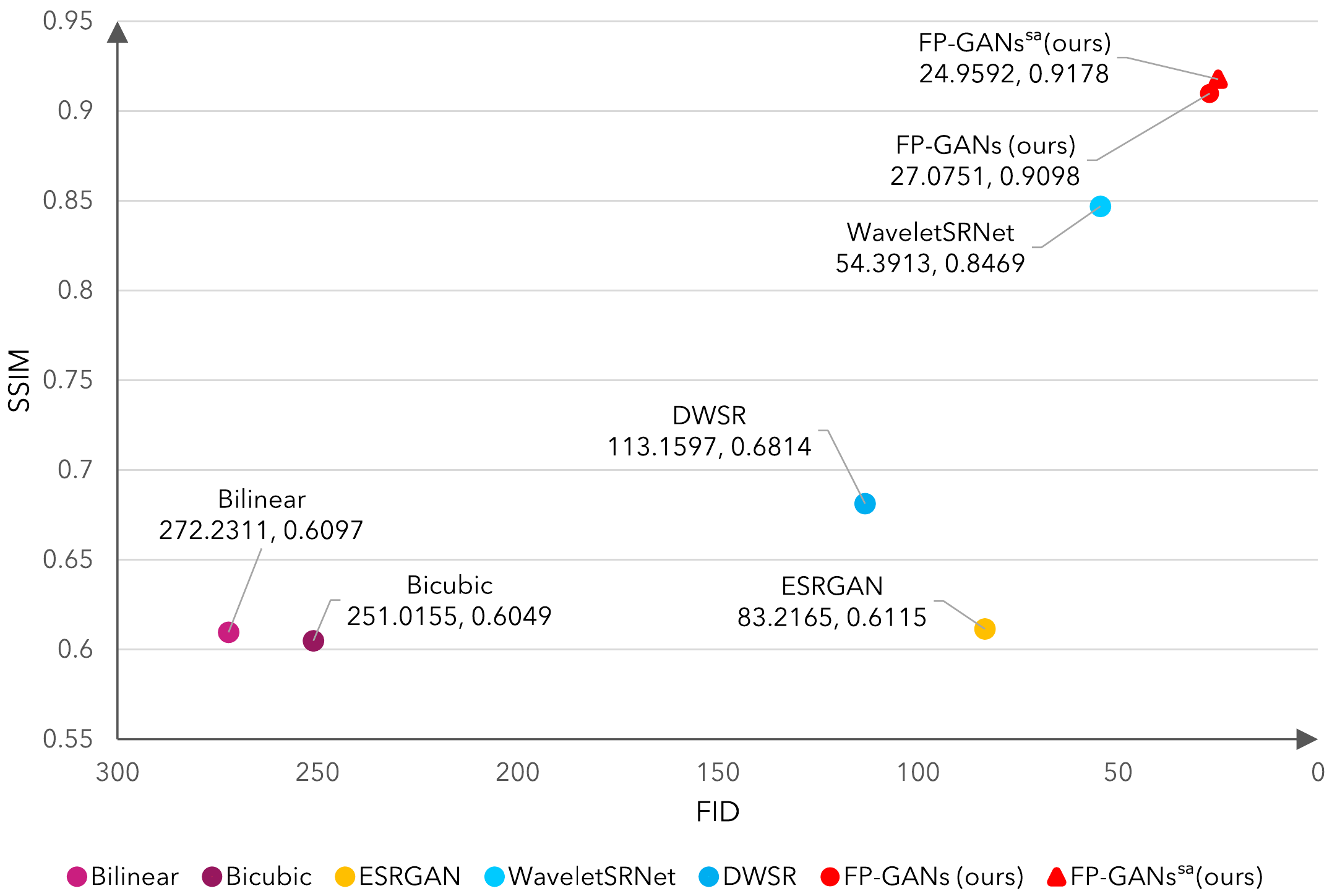}
	\caption{Comparison with different categories of super-resolution methods: 
	interpolation based methods (Bilinear, Bicubic), GAN based method (ESRGAN), 
	wavelet based methods (WaveletSRNet, DWSR) on SSIM and FID metrics. 
	The SSIM and FID values are evaluated on MultiRes\_7T dataset with scale factor $\times4$.}
	\label{result_scatter}
\end{figure}

\begin{figure*}[htb]
    \centering
      \includegraphics[width=\textwidth]{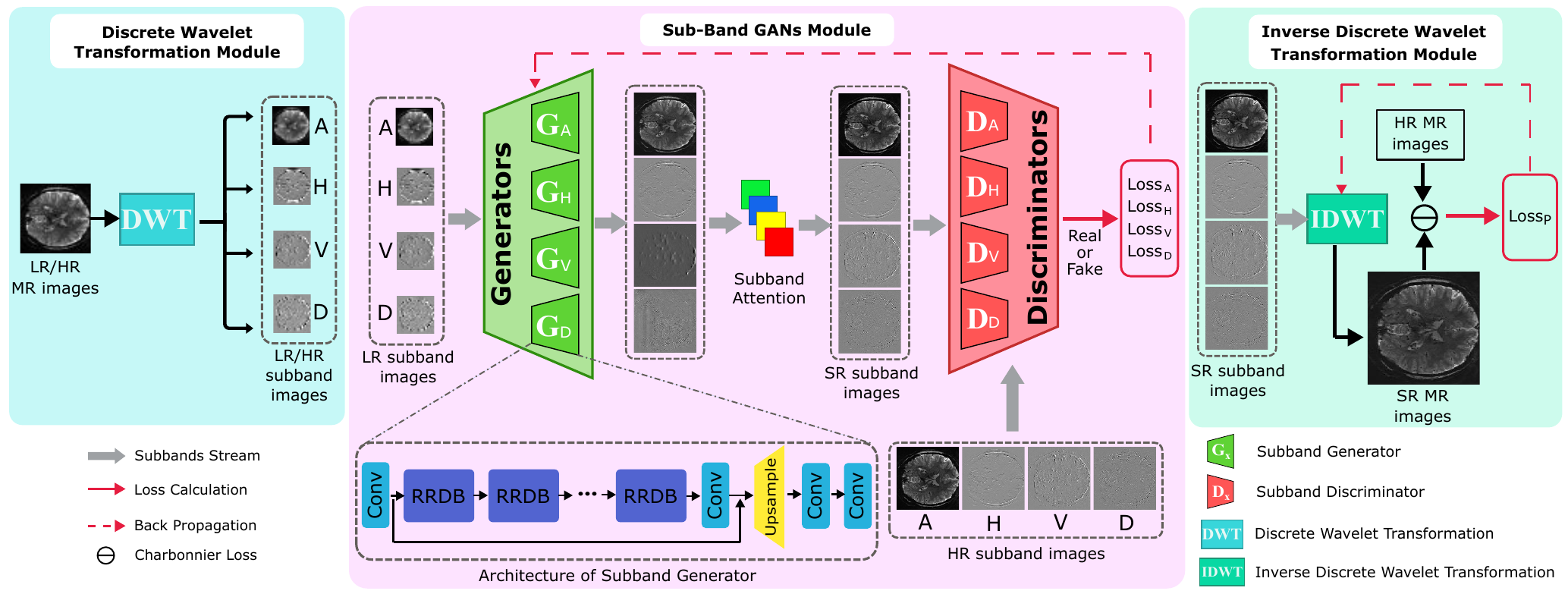}
	  \caption{The architecture of FP-GANs. A, H, V and D represent the approximate, 
	  vertical, horizontal and diagonal information of MR images respectively. 
	  $(G_A, D_A), (G_H, D_H), (G_V, D_V), (G_D, D_D)$ are sub-band GANs that learn 
      and predict higher resolution wavelet coefficients.}
    \label{FPGANs_framework}
\end{figure*}

Since the success of SRCNN\cite{25_dong2014learning, 19_dong2015image}, single image 
super-resolution (SISR) has achieved significant improvement qualitatively and quantitatively 
in recent years. It has attracted lots of interests in MR image super-resolution.  
For example, Cherukuri\cite{50_cherukuri2019deep} enhanced deep MR image super-resolution by 
a deep network that exploits a low-rank structure and a sharpness prior of MR images. 
Lyu\cite{51_lyu2020mri} presented an ensemble learning framework based on multiple GANs, 
where the complementary priors of five super-resolution algorithms 
(i.e. ZIP, BI, NEDI, SC and A+) are combined into a GAN model. 
However, most of existing deep learning based SISR models purely apply CNN for features 
extraction. It constrains the anatomical detail sensitivity of model and eventually leads 
to over-smooth result.





In this paper, a novel super-resolution model---fine structure perceptive generative 
adversarial networks, called fine perceptive GANs (FP-GANs), is designed to alleviate   
the detail insensitive problem of conventional CNN based models. 
As shown in Fig. \ref{FPGANs_framework}, rather than mapping low-resolution MR images to 
high-resolution counterparts directly, FP-GANs conducts super-resolution in a 
divide-and-conquer manner with multiple wavelet based generative adversarial networks.

More specifically, one MR image is \textbf{firstly} decomposed into four sub-band images 
(i.e. LL, LH, HL, HH) in wavelet domain by discrete wavelet transformation (DWT). 
The four sub-band images represent low-frequency global topology and high-frequency textures 
respectively. 
\textbf{Then}, for each sub-band image, a sub-band generative adversarial 
network (sub-band GAN) is deployed to learn the mapping from LR sub-band image to the 
corresponding HR sub-band image. With the scheme that learning the distribution of HR 
sub-bands respectively, it simplifies the mapping task from LR sub-bands to HR sub-bands 
and thus stabilizes the training procedure of GAN. 
What's more, dealing with global topology and detailed texture separately encourage 
the proposed model to be finer structure sensitive than the traditional deep CNN models. 
\textbf{Next}, the generated HR sub-band images are weighted by sub-band attention 
so that the model is able to adjust attention on each sub-band adaptively. 
\textbf{Finally}, the weighted SR sub-band images are inversely reconstructed into a SR MR 
image through inverse discrete wavelet transformation (IDWT). IDWT is implemented to be 
optimizable\cite{17_cotter2019learnable} and hence greatly improves the quality of SR result.

As illustrated in Fig. \ref{result_scatter}, FP-GANs outperforms the competing models with 
higher structural similarity index (SSIM)\cite{24_wang2004image} and lower 
Fr\'echet Inception Distance (FID)\cite{23_DOWSON1982450}. 
The main contributions of this paper are summarized as follows: 

1) A fine perceptive framework that exploits GANs to process different kinds of texture 
separately in wavelet domain is proposed. In this manner, it can effectively alleviate the 
detail insensitive problem. 
Besides, it simplifies the SR task for a single GAN, which accelerates and stabilizes 
distribution learning to some extent.

2) Unlike channel attention that trades off attention on feature channels, in this work, 
sub-band attention is designed to allocate attention on sub-band images adaptively. 
Extensive experiments demonstrate that sub-band attention contributes to balancing different 
kinds of textures.

3) Inverse discrete wavelet transformation is implemented optimizable in FP-GANs.
It facilitates factoring in the structure of the whole recomposed MR image and hence greatly 
improves the super-resolution performance.










\section{Related work}

\subsection{Image Reconstruction in Wavelet Domain}

There has been several studies exploring image reconstruction in wavelet domain due to the 
time-frequency localization of wavelet transformation. 
For example, Bae\cite{26_bae2017beyond}, Huang\cite{12_huang2017wavelet} and 
Guo\cite{13_guo2017deep} designed a deep convolutional neural network (CNN) to reconstruct 
high-resolution images with low-resolution wavelet sub-bands for capturing high-frequency 
textures. 
However, as the influence of high-frequency textures is relatively small, 
the textures still would be ignored if the wavelet sub-bands were processed jointly 
in one CNN model. 
Liu\cite{14_liu2018multi} exploited a multi-level wavelet CNN (MWCNN) to trade off 
receptive field size and computational efficiency, where wavelet transformation was 
introduced to reduce the size of feature map. 
Li\cite{16_li2019global} proposed a wavelet-domain global and local consistent age 
generative adversarial network (WaveletGLCA-GAN) to synthesize faces conditioned on age 
in frequency domain. 
To achieve better trade-off between objective and perceptual quality, 
Deng\cite{47_deng2019wavelet} divided the objective quality affected elements from 
perceptual ones with stationary wavelet decomposition and conquered them separately for  
different target. 
Xiao\cite{44_xiao2020invertible} proposed invertible rescaling net (IRN), 
an invertible bijective transformation based on wavelet transformation, 
to rescale the down-sampling images losslessly. 
Zhang\cite{53_zhangtexture} utilized wavelet texture loss with the objective to enhance 
more high-frequency components.

\subsection{Attention Mechanism}

Attention can be defined as a mechanism that allocating more limited available 
processing resources to more informative components of an object\cite{29_hu2018squeeze}.
Recently, attention mechanism (e.g. spatial attention, self attention and channel attention) 
has been applied to deep neural networks in many 
studies\cite{34_chen2017sca, 31_zhang2018image, 32_li2018tell}. 
Hu\cite{29_hu2018squeeze} proposed squeeze-and-excitation (SE) block to model 
channel-wise relationships to obtain significant performance improvement for image 
classification. Wang\cite{33_wang2017residual} proposed residual attention network 
for image classification with a trunk-and-mask attention mechanism. 
Zhang\cite{31_zhang2018image} utilized channel attention mechanism to adaptively rescale 
channel-wise features maps by considering relevance among channels for highly accurate 
image super-resolution. 
Bastidas\cite{35_bastidas2019channel} applied soft attention on individual channels 
for semantic segmentation.

\subsection{Network architecture}

As shown in Fig. \ref{FPGANs_framework}, FP-GANs consists of 3 parts in general. 
Let $I_{LR}, I_{HR}, I_{SR}$ represent low-resolution, high-resolution and super-resolution 
MR images respectively. $\left\{LR, HR\right\}$ and $\left\{A, H, V, D\right\}$ 
are optional list, where A, H, V, D represent the approximation, vertical, horizontal and 
diagonal texture of magnetic resonance (MR) image respectively. 
The processing procedure in FP-GANs can be summarized and formulated as follows:

1) Wavelet transformation module. Discrete wavelet transformation (DWT) is applied to 
decompose MR images into four sub-bands (i.e. LL, LH, HL, HH). These sub-bands contain  
low-frequency global information and high-frequency texture information in wavelet domain.

\begin{equation}
  \begin{split}
  &I_{A}^{\left\{LR, HR\right\}}, I_{H}^{\left\{LR, HR\right\}}, 
	I_{V}^{\left\{LR, HR\right\}}, I_{D}^{\left\{LR, HR\right\}} \\
	= & {\rm{DWT}} \Big( I^{\left\{LR, HR\right\}} \Big) .
  \end{split}
\end{equation}

2) Sub-band generative adversarial networks module. A group of generative adversarial 
networks (GANs) take low-resolution sub-band images as input and learn to generate the 
corresponding high-resolution sub-band images.

\begin{equation}
	I_{\left\{A, H, V, D\right\}}^{SR} 
	= G_{\left\{A, H, V, D\right\}} \Big( I_{\left\{A, H, V, D\right\}}^{LR} \Big) .
\end{equation}

Between the generator and discriminator of sub-band GANs, sub-band attention learns to 
trade off the approximation, horizontal, vertical and diagonal textures of the SR sub-band 
images. It enhances visual quality by balancing the global topology and detailed textures. 
The detail of sub-band attention can be found in supplementary material.

\begin{small}
\begin{equation}
	\Big[ {I_{A}^{SR}}_{w}, {I_{H}^{SR}}_{w}, {I_{V}^{SR}}_{w}, {I_{D}^{SR}}_{w} \Big] 
	= W \cdot \Big[ I_{A}^{SR}, I_{H}^{SR}, I_{V}^{SR}, I_{D}^{SR} \Big] .
\end{equation}
\end{small}

3) Inverse wavelet transformation module. Inverse discrete wavelet transformation 
reconstructs SR MR images by synthesizing the weighted SR sub-band images.

\begin{equation}
	I^{SR} = {\rm{IDWT}} 
	\Big( {I_{A}^{SR}}_{w}, {I_{H}^{SR}}_{w}, {I_{V}^{SR}}_{w}, {I_{D}^{SR}}_{w} \Big) .
\end{equation}

\subsection{Discrete wavelet transformation}

\begin{figure}[htb]
	\centering
	\subfloat[]
	{\includegraphics[width=0.45\textwidth]{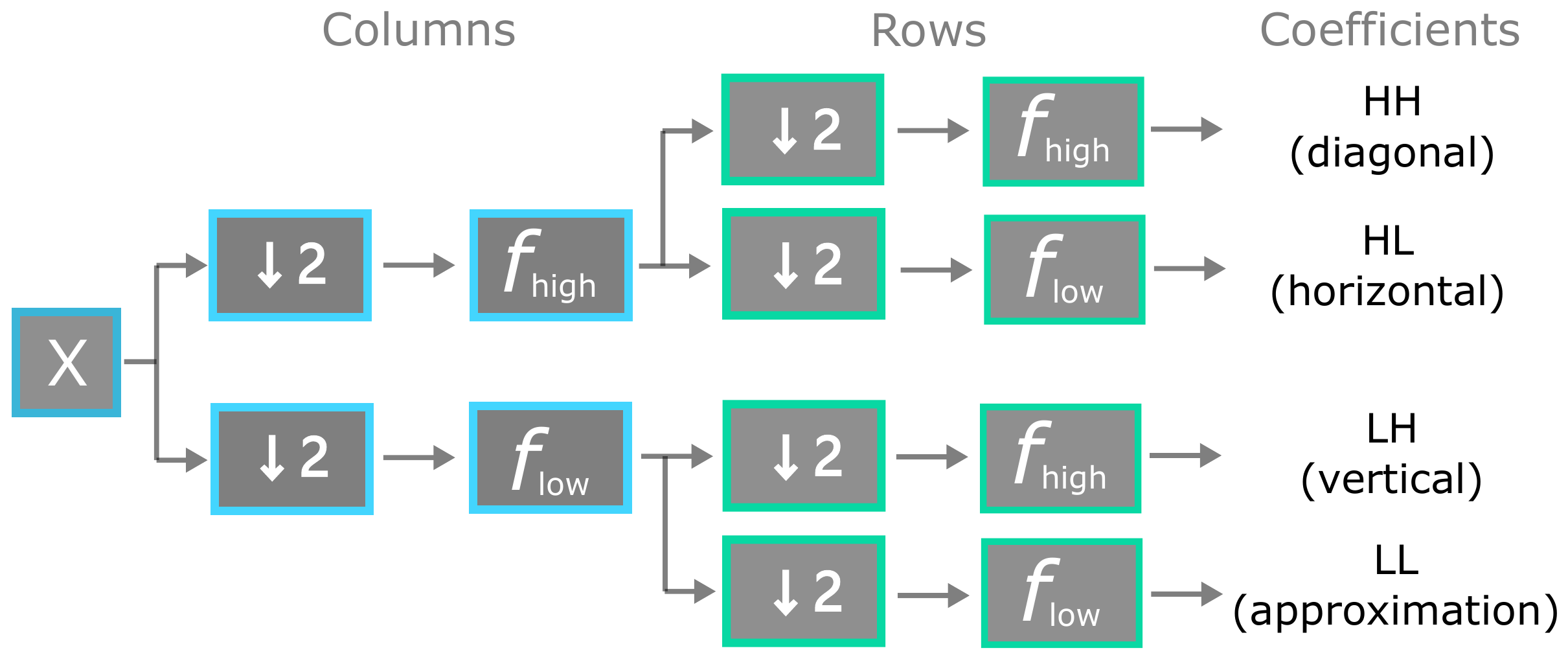}%
	\label{2D_DWT}}
	\\
	\subfloat[]
	{\includegraphics[width=0.45\textwidth]{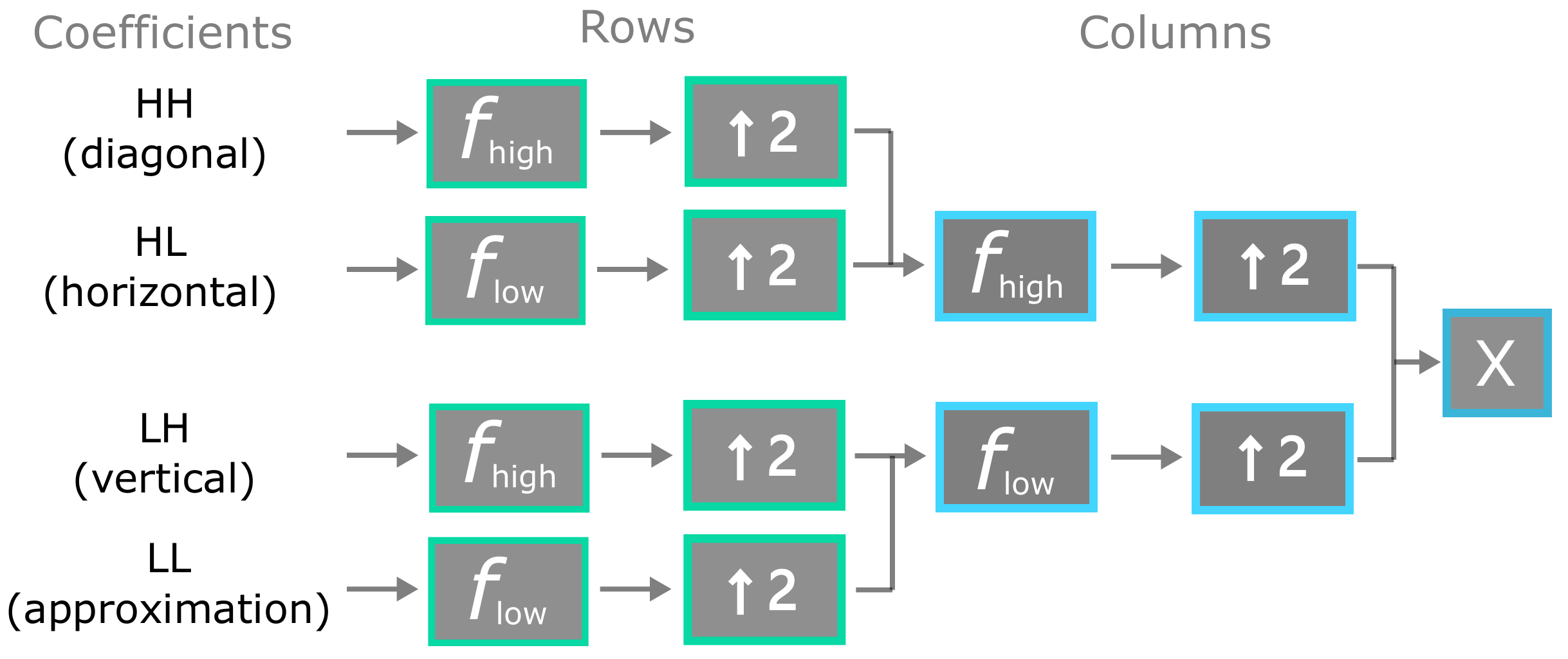}%
	\label{2D_IDWT}}
\caption{Procedure of wavelet transformation.
(a) 1-level wavelet transformation, decomposing an image into four sub-bands (i.e. LL, LH, HL, HH). 
(b) 1-level wavelet transformation, recomposing the sub-bands into original image. 
$f_{high}$ and $f_{low}$ represent high-pass filter and low-pass filter respectively. }
\label{2D_transformation_procedure}
\end{figure}

1-level discrete wavelet transformation (DWT) decomposes an image into four smaller 
2D sub-bands: LL (global approximation), LH (horizontal texture), HL (vertical texture), 
HH (diagonal texture), as Fig. \ref{2D_DWT} shows. The sub-band images can be losslessly  
recomposed into the original MR image by inverse discrete wavelet transformation (IDWT) 
as shown in Fig. \ref{2D_IDWT}. 

\begin{figure}[htb]
	\centering
	\includegraphics[width=0.48\textwidth]{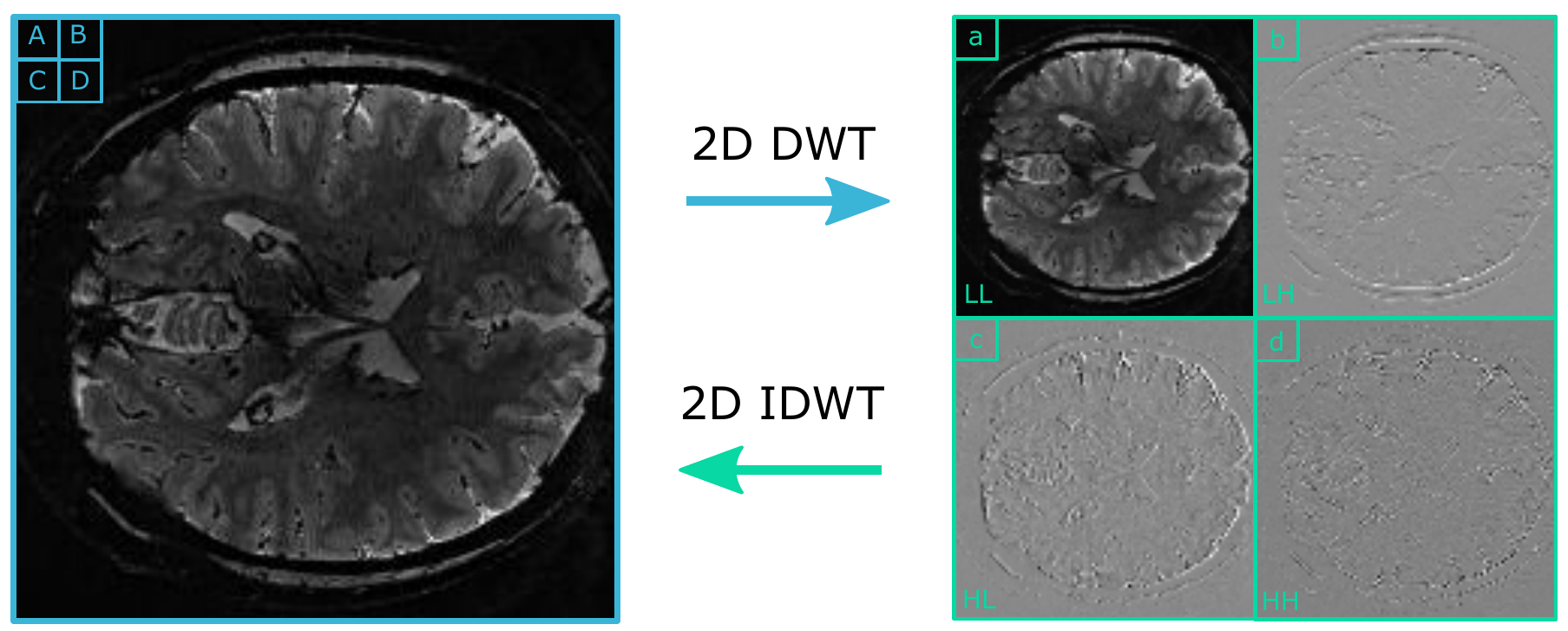}
	\caption{Instance of Haar based discrete wavelet transformation (DWT) and 
	inverse discrete wavelet transformation (IDWT).}
	\label{DWT_IDWT_sample}
\end{figure}

Inspired by the lossless recomposition procedure, 1-level 2D DWT is employed in FP-GANs 
to divide different oriented textures and conquer them. 
Take Haar based DWT as an example, the transformation instance can be illustrated as 
Fig. \ref{DWT_IDWT_sample}.

Let the filter of Haar based wavelet transformation defined as Equation (\ref{haar_filter}), 
then the transformation procedure of Fig. \ref{DWT_IDWT_sample} can be concisely computed 
as Equation (\ref{haar_DWT_IDWT}).

\begin{equation}
	\begin{split}
	f_{LL}=\
	  \begin{bmatrix}
		  \makebox[\widthof{$-1$}]{1} & \makebox[\widthof{$-1$}]{1}\\
		  1 & 1
	  \end{bmatrix} 
	  \hspace{8pt}
	&f_{LH}=\
	  \begin{bmatrix}      
		  -1 & -1\\
		  1 & 1
	  \end{bmatrix}
	  \hspace{8pt}\\
	f_{HL}=\
	  \begin{bmatrix}
		  -1 & \makebox[\widthof{$-1$}]{1}\\
		  -1 & 1
	  \end{bmatrix}
	  \hspace{8pt}
	&f_{HH}=\
	  \begin{bmatrix}
		  \makebox[\widthof{$-1$}]{1} & -1\\
		  -1 & 1
	  \end{bmatrix}
	  \hspace{8pt}
	\end{split},
\label{haar_filter}
\end{equation}

\begin{small}
\begin{equation}
	\begin{cases}
		a=A+B+C+D \\  
 		b=A+B-C-D \\
 		c=A-B+C-D \\ 
 		d=A-B-C+D \\
	\end{cases} \\
	\rightleftharpoons
	\hspace{8pt}
	\begin{cases}
		A=(a+b+c+d)/{4} \\  
		B=(a+b-c-d)/{4} \\
		C=(a-b+c-d)/{4} \\ 
		D=(a-b-c+d)/{4} \\
	\end{cases}.
\label{haar_DWT_IDWT}
\end{equation}
\end{small}where $A, B, C, D$ represent the pixel intensity in the original MR image in 
Fig. \ref{DWT_IDWT_sample}, and $a, b, c, d$ represent the pixel intensity in four 
sub-bands respectively.

\subsection{Sub-band GANs}

The second part of FP-GANs consists of four pairs of generative adversarial network 
with same architecture, namely sub-band GANs. Each GAN focuses on one sub-band  
(i.e. LL, LH, HL, HH), learning to generate higher resolution sub-band images from 
low-resolution ones. The generators are trained to capture the distribution of HR sub-band 
images, while the discriminators are encouraged to estimate whether the images are real 
or generated. 

Suppose $X$ denotes low-resolution MR images dataset and Y denotes high-resolution MR images 
dataset as ground truth. Let $\{x_i\}_i^N\subset X$ and $\{y_i\}_i^N\subset Y$, $N$ denotes 
the number of MR images. Then the adversarial procedure of traditional super-resolution 
GAN can be formulated as following min-max problem:

\begin{small}
\begin{equation}
	\mathop{min} \limits_{G} \mathop{max} \limits_{D} 
	\mathbb{E}_{y \in Y} \Big[\log \big(D(y) \big) \Big]
	+ \mathbb{E}_{x \in X} \Big[1 - \log{ \Big(D \big(G(x) \big) \Big)} \Big] .
\end{equation}
\end{small}

Due to the usage of relativistic discrimination regime, the objective function in FP-GANs 
can be formulated as:

\begin{small}
\begin{equation}
	\mathop{min} \limits_{G} \mathop{max} \limits_{D}
	\mathbb{E}_{y \in Y} \Big[\log \Big(D \big(y, G(x) \big) \Big) \Big]
	+ \mathbb{E}_{x \in X} \Big[\log \Big(1 - D \big( G(x), y \big) \Big) \Big] ,
\end{equation}
\end{small}where 
\begin{small} $D\big(y, G(x)\big)= {\rm sigmoid} \Big( \mathcal{D}(y)-\mathbb{E}_{x \in X}
\big[ \mathcal{D} \big( G(x) \big) \big] \Big)$\end{small}, 
$\mathcal{D}(x)$ denotes the output of discriminator.

\subsubsection{Generator}

\begin{figure}[htb]
  \centering
  \includegraphics[width=0.45\textwidth]{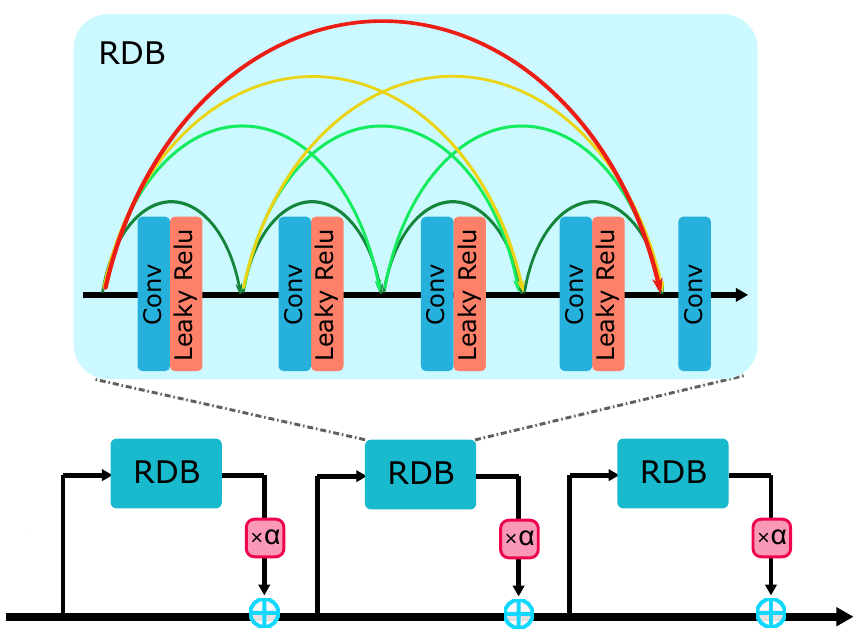}
	\caption{The structure of residual-in-residual-dense-block (RRDB) in generator.} 
	\label{RRDB_architecture}
\end{figure}

As illustrated in Fig. \ref{FPGANs_framework}, the generator of sub-band GAN consists of 
5 parts: first convolutional layer, residual-in-residual-dense-blocks (RRDBs) module 
(Fig. \ref{RRDB_architecture}), 
trunk convolutional layer, upsampling module and final convolutional layer referred as 
ESRGAN\cite{18_wang2018esrgan} and RFB-ESRGAN\cite{41_zhang2020ntire}. 

\begin{figure*}[htb]
	\centering
	\includegraphics[width=0.65\textwidth]{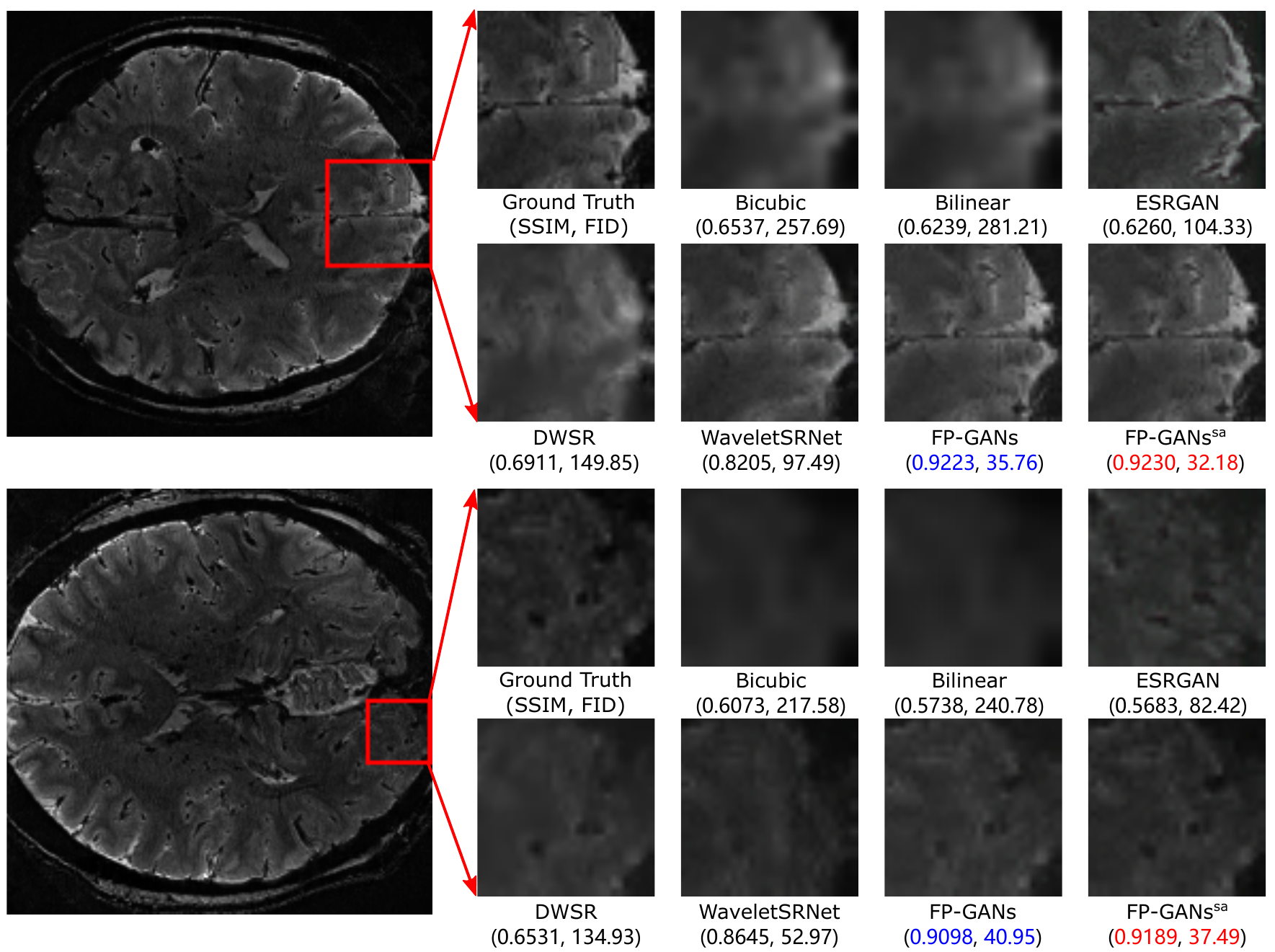}
	\caption{Visual comparison of detail recovery performance on MultiRes\_7T dataset. 
	The images are gained with super-resolution scale factor $\times 4$. 
	Red and blue indicate the best and the second best performance, respectively.}
	\label{single_detail}
\end{figure*}

\begin{figure*}[htb]
	\centering
	\includegraphics[width=0.95\textwidth]{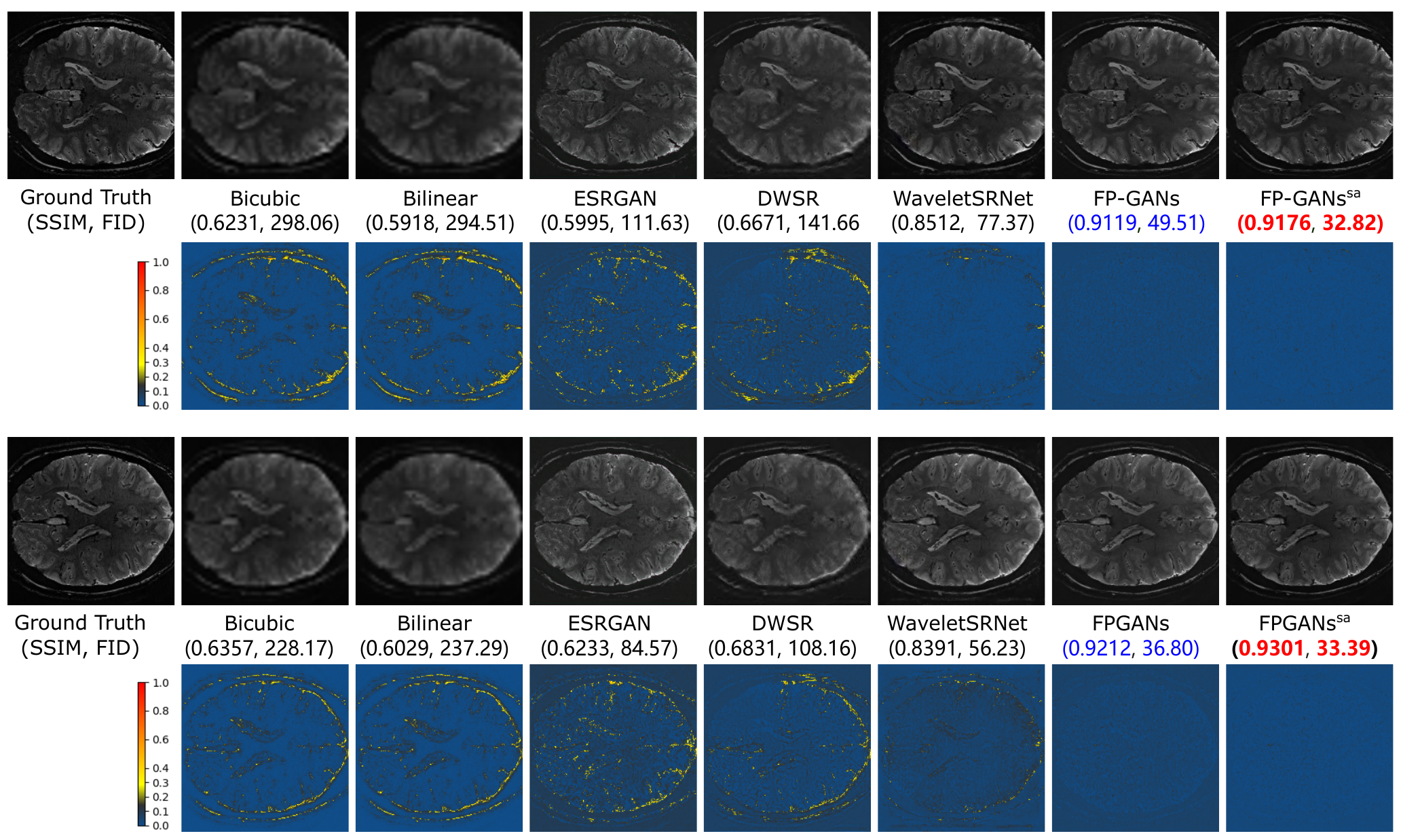}
	\caption{Difference comparison among Bicubic, Bilinear, ESRGAN, DWSR, WaveletSRNet, 
	FP-GANs and FP-GANs$^{sa}$ with heatmap. 
	Brighter pixel means more difference to the ground truth.}
	\label{Error_map}
\end{figure*}

\subsection{Training Loss}

Adversarial loss, wavelet loss and pixel loss are combined to guide the training of FP-GANs 
from difference perspective. 

\subsubsection{Adversarial Loss}
With prior knowledge that real and generated MR images account for a half 
respectively while supervisedly training, adversarial loss based on relativistic 
GAN\cite{9_jolicoeur2018relativistic} is introduced to stabilize the training and improve 
the quality of the generated MR image. The adversarial loss function of four sub-band 
GANs $(G_A, G_H, G_V, G_D)$ can be defined as:

\begin{small}
\begin{equation}
  \label{1_equation}
  \begin{split}
  Loss_{D_{\left\{ A,H,V,D \right\}}} 
	= &- \mathbb{E}_{y \in Y} \Big[\log \big( D_{\left\{ A,H,V,D \right\}}(y, x_f) \big) \Big] \\
	&- \mathbb{E}_{x \in X} \Big[\log  \big(1-D_{\left\{ A,H,V,D \right\}}(y, x_f) \big) \Big] ,
  \end{split}
\end{equation}
\end{small}
\begin{small}
\begin{equation}
  \label{2_equation}
  \begin{split}
	Loss_{G_{\left\{ A,H,V,D \right\}}} 
	= &- \mathbb{E}_{y \in Y} \Big[ \log \big( 1-D_{\left\{ A,H,V,D \right\}}(y, x_f) \big) \Big]\\
  &- \mathbb{E}_{x \in X} \Big[ \log \big(D_{\left\{ A,H,V,D \right\}}(y, x_f) \big) \Big] ,
  \end{split}
\end{equation}
\end{small}where 
$D(y, x_f) = {\rm sigmoid} \Big(\mathcal{D}(y) - \mathbb{E}_{x \in X} \big[ \mathcal{D}(x_f) \big] \Big)$, 
$\mathcal{D}(x)$ denotes the output of discriminator, $x_f=G(x)$, and $\left\{ A,H,V,D \right\}$ 
is an optional list.

\subsubsection{Wavelet Loss}

As textures mainly are comprised of high-frequency components, 
wavelet loss is introduced for paying more emphasis on high frequency components. 
The contribution that eliminates over-smooth result and recovers more anatomical detail 
of MR image will be further demonstrated latter. 
Wavelet loss on each sub-band can be computed as:

\begin{small}
\begin{equation}
  \label{3_equation}
  \begin{split}
  &Loss_{W_{\{A,H,V,D\}}} \\
	= & \mathbb{E}_{x\in X,y\in Y} \Big[ \Big\| y_{\{A,H,V,D\}} - G_{\{A,V,H,D\}} \big(x_{\{A,H,V,D\}} \big) \Big\|_1 \Big] ,
  \end{split}
\end{equation}
\end{small}where $x_{\{A,H,V,D\}}$ and $y_{\{A,H,V,D\}}$ denote the low and high resolution 
sub-band images, respectively.

\subsubsection{Pixel Loss}
Pixel loss, factoring the global structure of the reconstructed image, 
is calculated by Charbonnier penalty function as Equation (\ref{4_equation}). 
Charbonnier penalty function is a differential variant of $L_1$ normalization. 
It can avoid the over-smooth effect of $L_2$ normalization. 

\begin{equation}
    \label{4_equation}
    \begin{split}
    &Loss_P \\
    = &\mathbb{E}_{x_f\in X,y\in Y} \Bigg[ \sqrt{ \big( y- {\rm{IDWT}}(x_f^A,x_f^H,x_f^V,x_f^D) \big)^{2} 
    + \epsilon^{2} } \Bigg] ,
    \end{split}
\end{equation}
where $x_f^A,x_f^H,x_f^V,x_f^D$ represent the sub-band images predicted by 
$G_{\{A,V,H,D\}}$ respectively. $\epsilon$ is a slack variable and empirically set as 
$1 \times 10^{-6}$.

\subsubsection{FP-GANs Loss}
Overall, the total loss function of generators in FP-GANs can be defined as:

\begin{equation}
  \label{5_equation}
  \begin{split}
  Loss_{G_{total}} 
  = &\lambda_1 Loss_A + \lambda_2 Loss_H + \lambda_3 Loss_V + \lambda_4 Loss_D \\
  &+ \beta Loss_P ,
  \end{split}
\end{equation}
where \begin{small}$Loss_{ \{A,H,V,D\} } = \alpha Loss_{ G_{\{A,H,V,D\}} } + 
Loss_{ W_{\{A,H,V,D\}} }$\end{small}, 
$\alpha$ is weighted value that trades off the adversarial loss and wavelet loss.
$\lambda_{1}, \lambda_{2}, \lambda_{3}$ and $\lambda_{4}$ are parameters that relatively 
control the penalty on corresponding sub-band and facilitate textures trade-off.

Analogously, the discriminators of FP-GANs can be optimized by minimizing the 
following function:

\begin{equation}
	\label{6_equation}
	Loss_{D_{total}} = Loss_{D_{A}} + Loss_{D_{H}} + Loss_{D_{V}} + Loss_{D_{D}} .
\end{equation}

\begin{figure*}[!ht]
	\centering
	\includegraphics[width=0.95\textwidth]{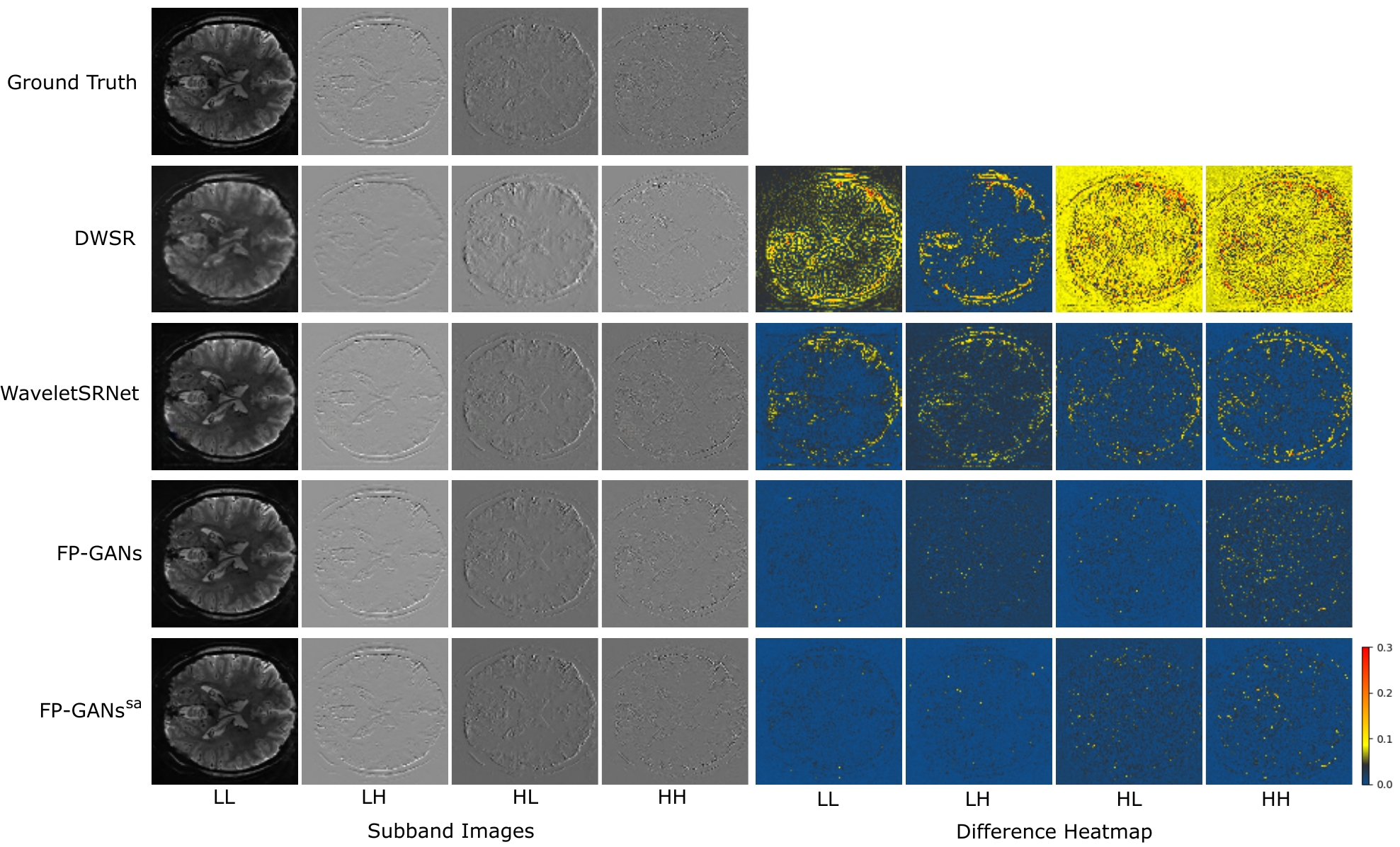}
	\caption{Comparison of sub-band recovery against the wavelet based methods 
	(i.e. DWSR, WaveletSRNet). Brighter pixel means more difference to the ground truth.}
	\label{sunband_error_map}
\end{figure*}

\renewcommand\arraystretch{2}
\begin{table*}[!ht]
	\centering
	\caption{Quantitative evaluation for super-resolution performance on MultiRes\_7T dataset}
	\label{tab1}
	\resizebox{0.8\linewidth}{!}{ %
	\begin{tabular}{ccccccccc}
		\toprule
		\textbf{Scale} & \diagbox{\textbf{Metrics}} {\textbf{Method}} & \textbf{Bicubic} & \textbf{Bilinear}& \textbf{ESRGAN} & \textbf{DWSR} & \textbf{WaveletSRNet} & \textbf{FP-GANs} & \textbf{{FP-GANs}$^{sa}$}\\
		\midrule
		\hline
		\multirow{3}{*}{$\times 2$}
		& \textbf{PSNR} &27.4957 &27.6820 &27.6074 &-- &24.7976 &\color{red}28.2176 &\color{blue}27.8190 \\
		& \textbf{SSIM} &0.8269 &0.7894 &0.8546 &-- &\color{red}0.9366 &\color{blue}0.9225 &0.9226 \\
		& \textbf{FID} &104.22 &124.53 &48.41 &-- &22.77 &\color{red}25.01 &\color{blue}26.01 \\
		\hline
		\hline
		\multirow{3}{*}{$\times 4$}
		& \textbf{PSNR} &25.2539 &24.8464 &21.8490 &23.2888 &\color{blue}26.6529 &25.0345 & \color{red}27.5964\\
		& \textbf{SSIM} &0.6408 &0.6097 &0.6115 &0.6800 &0.8469 &\color{blue}0.9098 & \color{red}0.9178\\
		& \textbf{FID} &251.01 &272.23 &83.21 &113.16 &54.39 &\color{blue}27.07 & \color{red}24.95\\
		\hline
		\hline
		\multirow{3}{*}{$\times 8$}
		& \textbf{PSNR} &\color{red}22.8815 &\color{blue}22.7049 &17.8975 &-- &21.7509 &20.7491 &21.2957\\
		& \textbf{SSIM} &0.4528 &0.4412 &0.3463 &-- &0.6595 &\color{blue}0.7947 &\color{red}0.8036\\
		& \textbf{FID} &318.74 &403.33 &\color{blue}98.03 &-- &\color{red}83.06 &124.71 &116.13\\
		\bottomrule
	\end{tabular}} %
\end{table*}

\begin{figure*}[htb]
	\centering
	\subfloat[]
	{\includegraphics[width=0.45\textwidth]{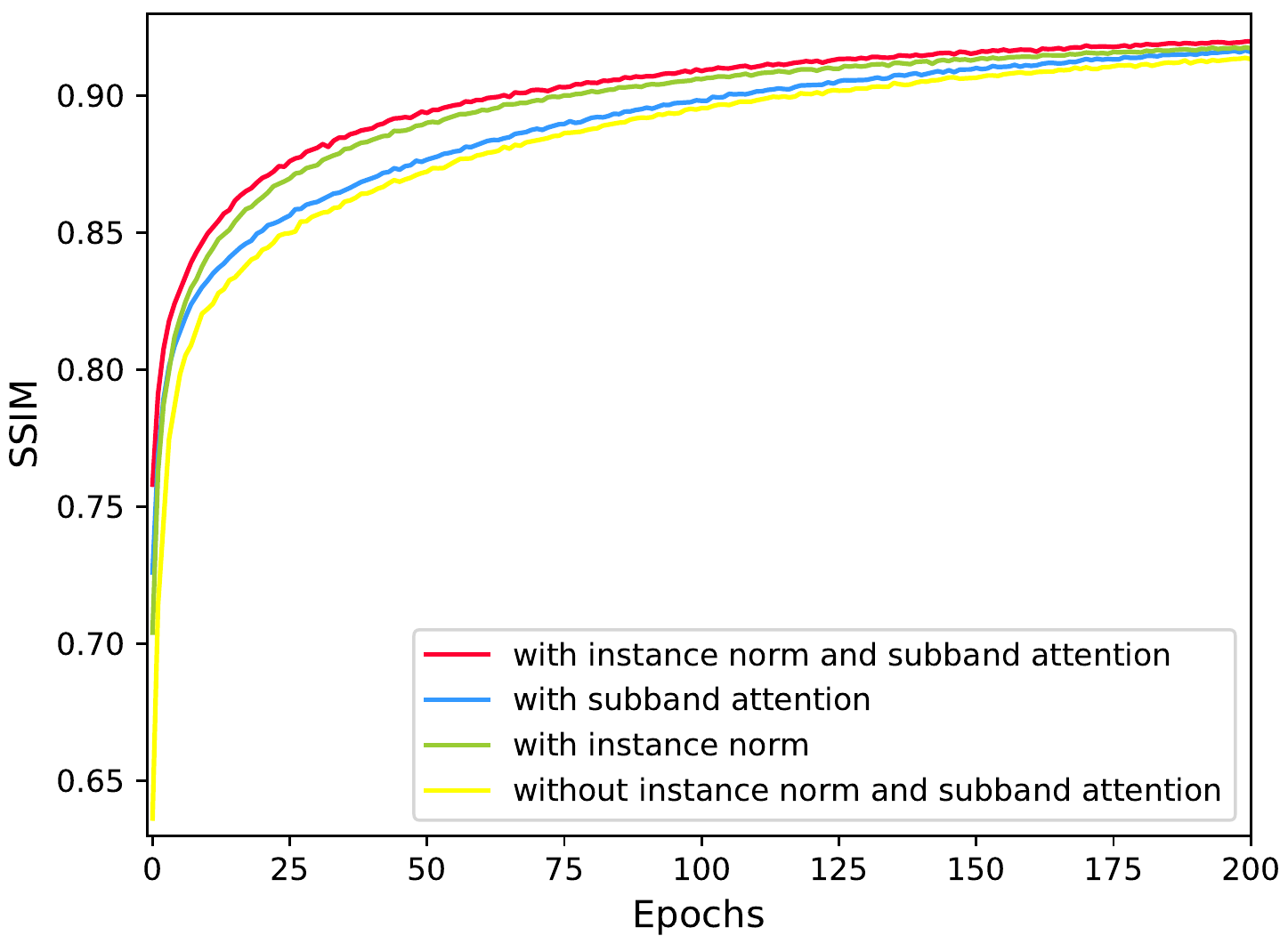}%
	\label{in_sa}}
	\subfloat[]
	{\includegraphics[width=0.45\textwidth]{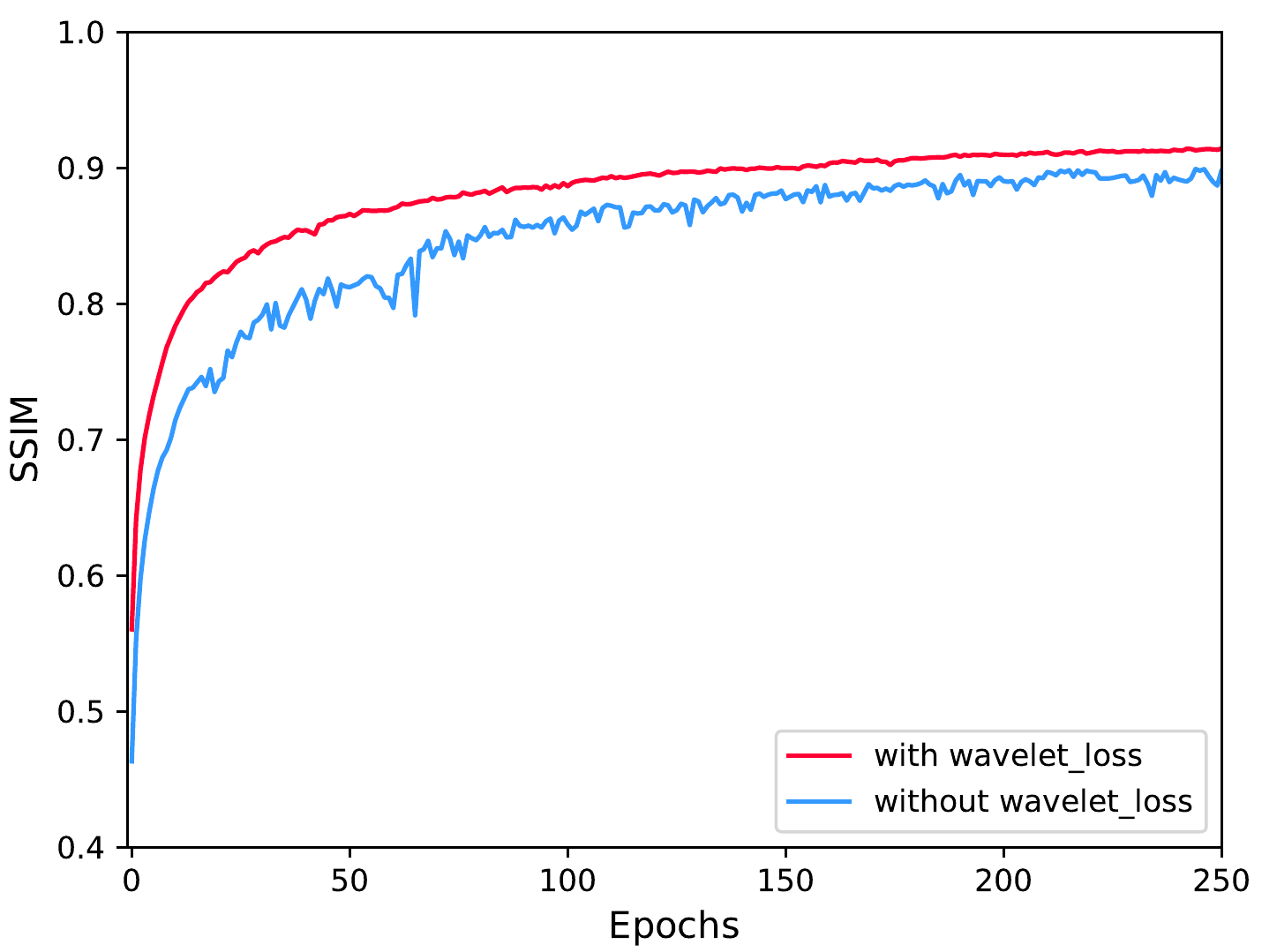}%
	\label{wavelet_loss}}
\caption{Ablation study: 
(a) Ablation study on instance normalization and sub-band attention. 
(b) Ablation study on wavelet loss.}
\label{subband_attention_weight}
\end{figure*}

\begin{figure*}[htb]
	\centering
	\subfloat[0th epoch]
	{\includegraphics[width=0.2\textwidth]{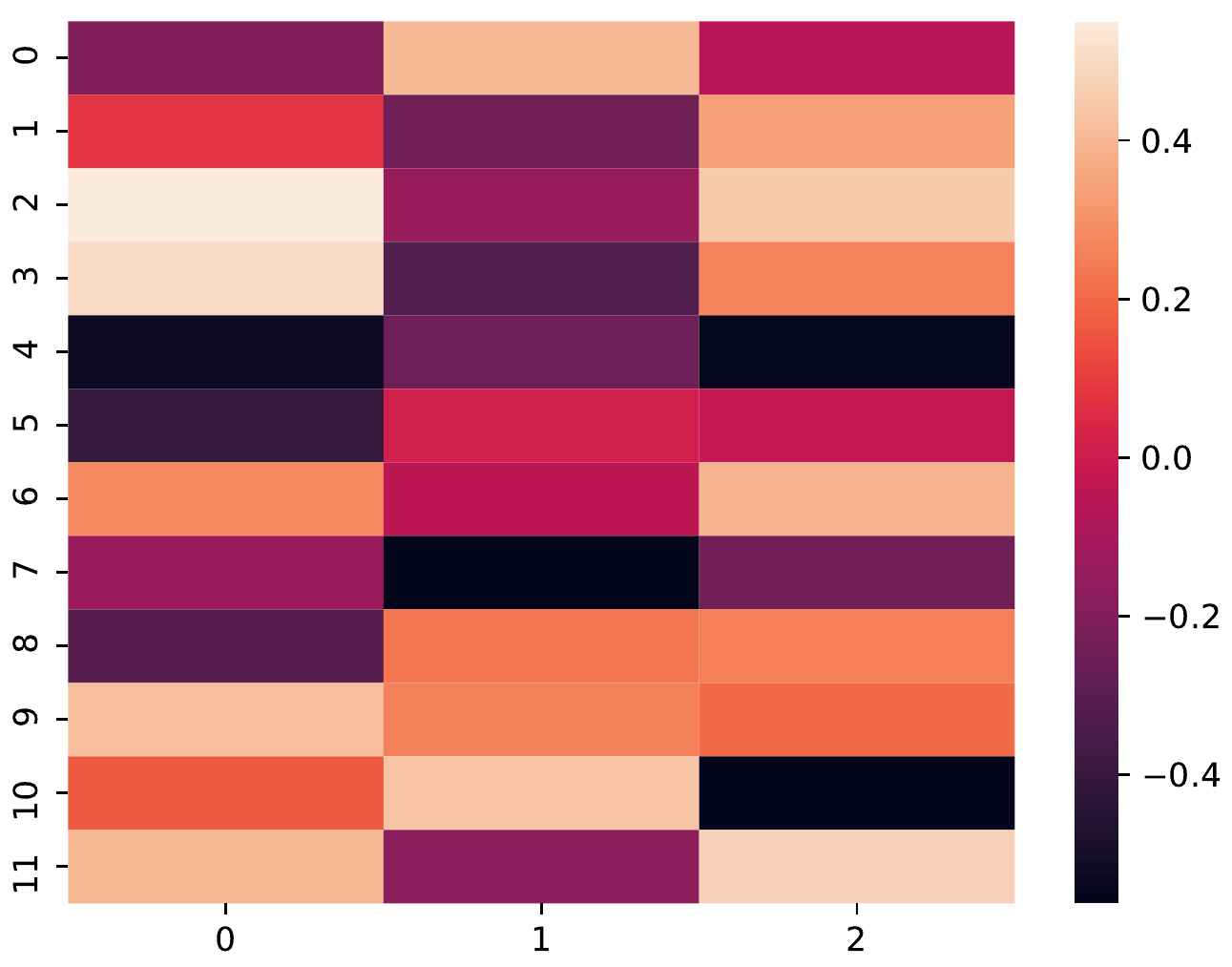}%
	\label{0epoch}}
	\subfloat[50th epoch]
	{\includegraphics[width=0.2\textwidth]{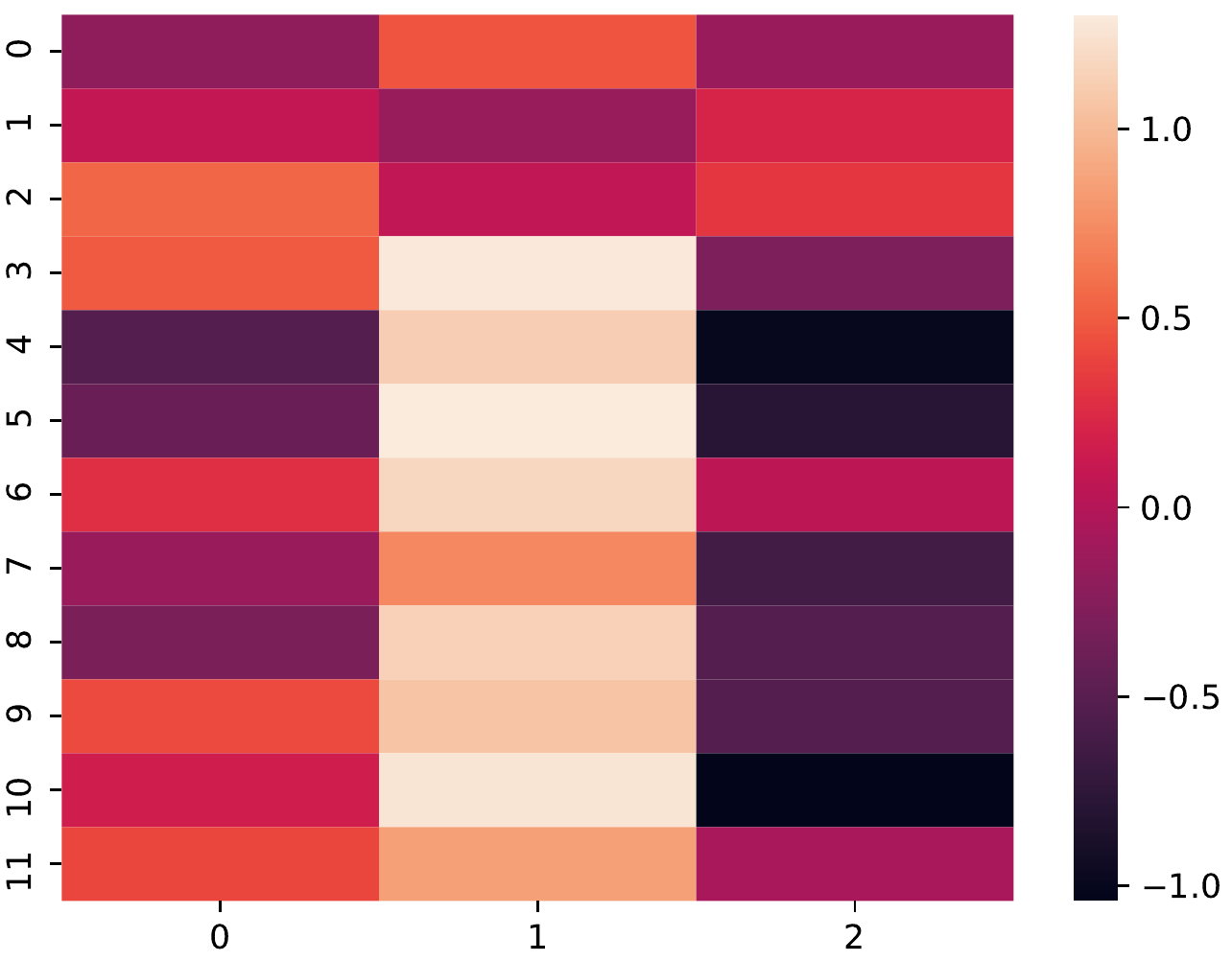}%
	\label{50epoch}}
	\subfloat[100th epoch]
	{\includegraphics[width=0.2\textwidth]{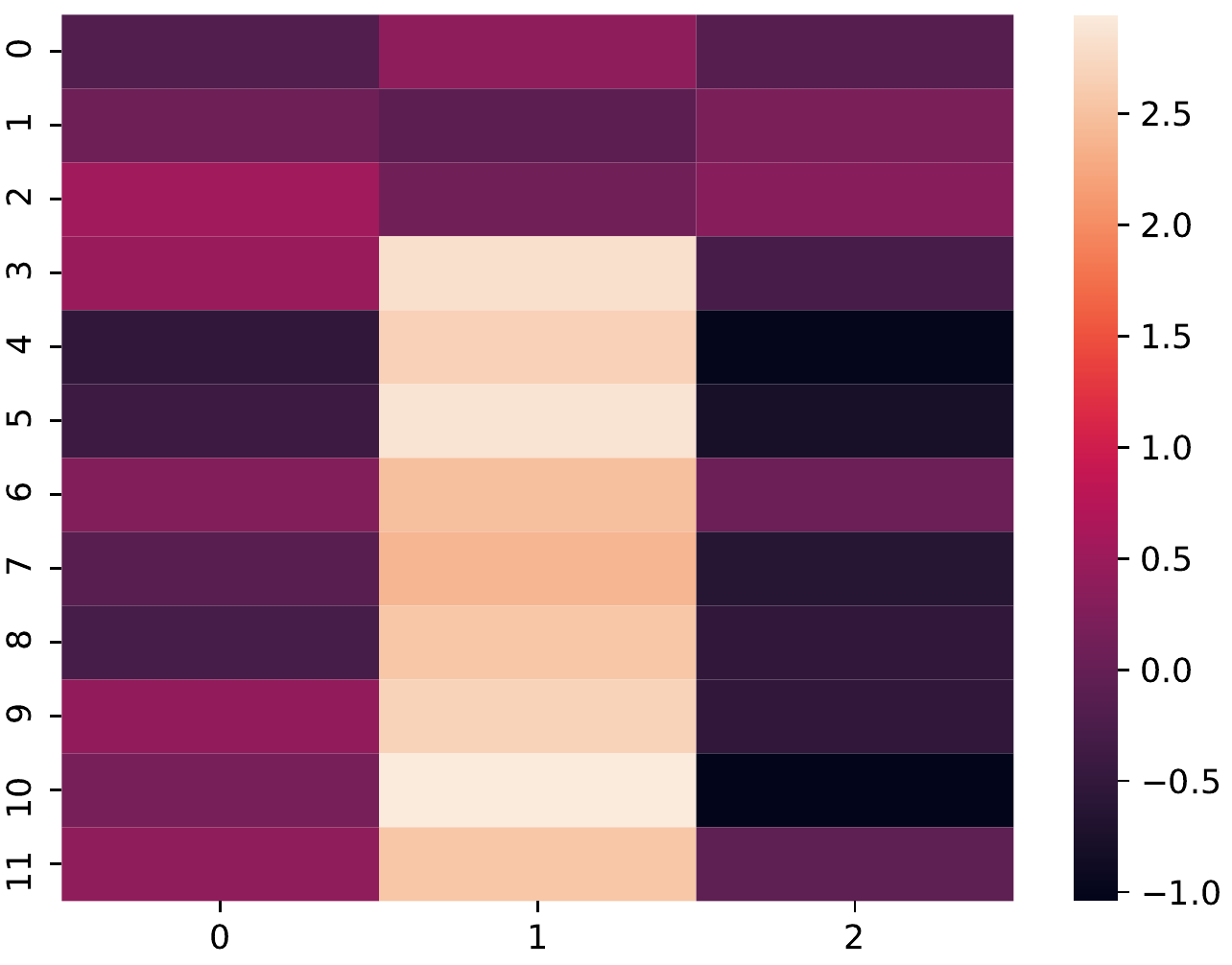}%
	\label{100epoch}}
	\subfloat[200th epoch]
	{\includegraphics[width=0.2\textwidth]{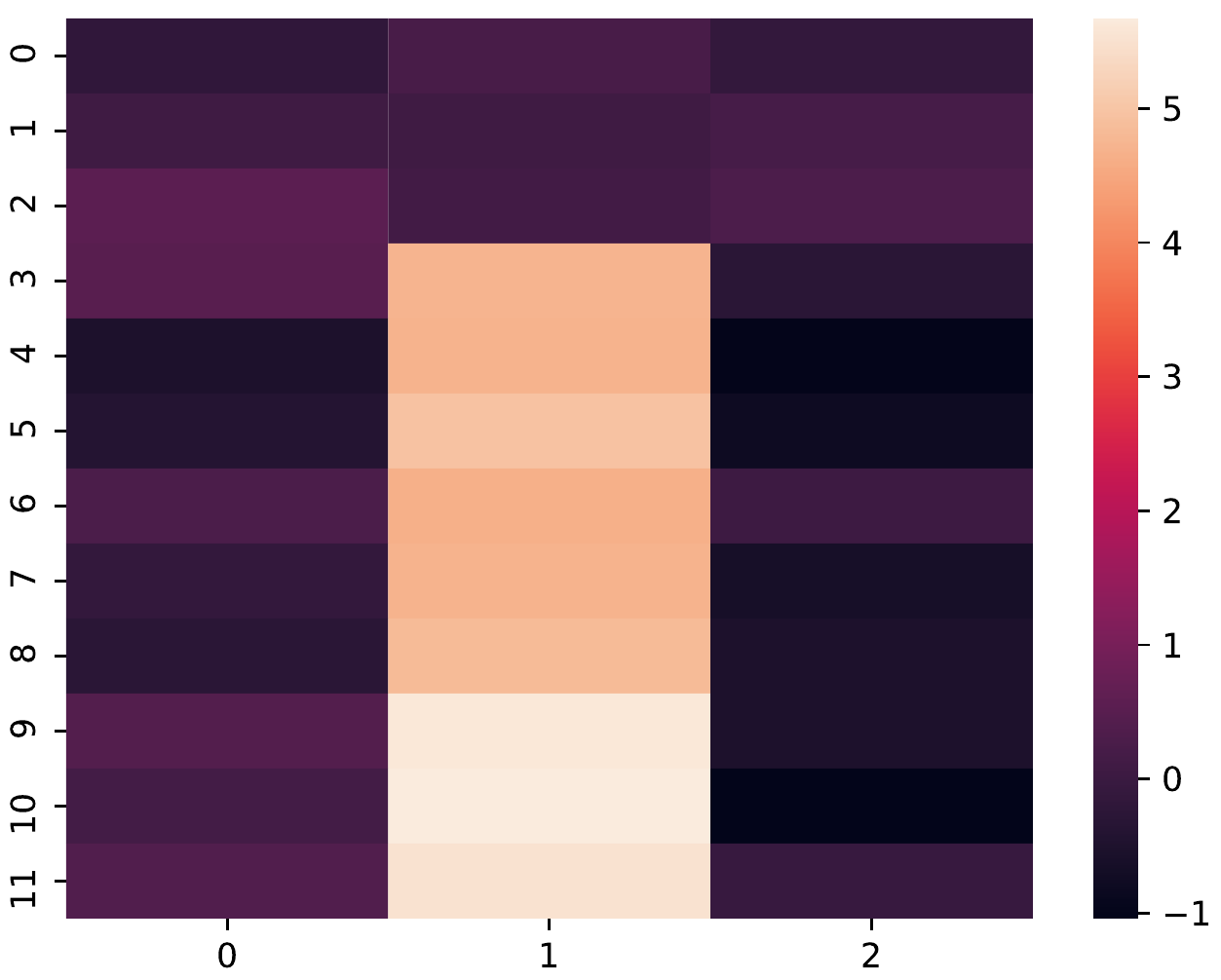}%
	\label{200epoch}}
	\subfloat[400th epoch]
	{\includegraphics[width=0.2\textwidth]{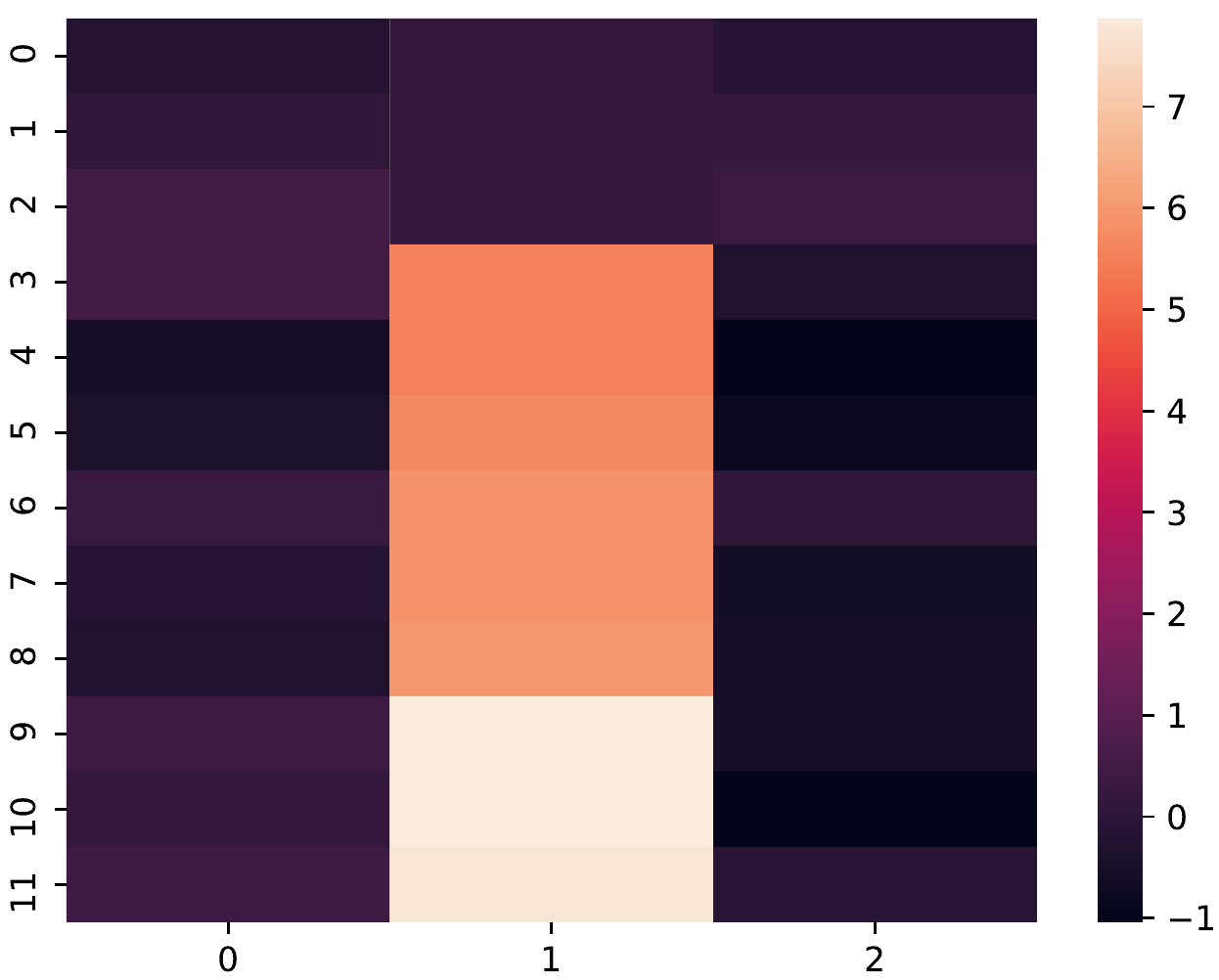}%
	\label{400epoch}}
\caption{The weight update procedure of sub-band attention (SA) module. 0-2 rows 
denote the weight of LL sub-band, and 3-5, 6-8, 9-11 rows denote LH, HL and HH 
sub-bands respectively.
(a) The weight of the initial stage, 0th epoch. (b), (c), (d), (e) are the weight of 
50th, 100th, 200th and 400th epoch.}
\label{sa_weight}
\end{figure*}

\section{Experiments}

\subsection{Data preparation}

Multi-resolution 7T (MultiRes\_7T) functional MR images data was acquired with 
Siemens 7T scanner from OpenfMRI database with accession number: 
ds000113c\cite{10_sengupta2017ultra}. 
The dataset consists of ultra high-field functional MR images recorded 
at four spatial resolutions (0.8 mm, 1.4 mm, 2.0 mm and 3.0 mm isotropic voxel size) from 
7 participants. 

In the experiment, 0.8 mm and 3.0 mm voxel size functional MR images with shape of 
$250\times 250\times 57$ and $66\times 66\times 37$ respectively are utilized.  
Firstly, slice the images into $(250\times 250$ and $66\times 66)$ along z-axis. 
Secondly, conduct center crop on the slices to cut off redundant content, 
and get corresponding grayscale images $(200\times 200)$ and $50\times 50)$. 
Thirdly, downsample the high-resolution images from $200\times 200$ with scale factor 
(i.e. 2, 4, 8) to acquire corresponding low-resolution images in size 
$(100 \times 100, 50 \times 50, 25 \times 25)$ and thus build the paired dataset. 

\subsection{Implementation and Training detail}

FP-GANs is implemented with Pytorch and the experiments are conducted on Nvidia Titan RTX. 
In generator, the number of residual in residual dense block (RRDB) is chosen as 16.
In discriminator, there are 10 convolutional layers followed by 2 full 
connected layers. 
FP-GANs is optimized with Adam\cite{11_kingma2014adam}. The batch size is set as 16. 
The learning rate of the generators is set as $1 \times 10^{-4}$ and the learning rate of 
discriminators is set as $5 \times 10^{-4}$. 
Besides, the learning rate of IDWT is set as $1 \times 10^{-4}$. 
Hyper-parameters $\lambda_1,\lambda_2,\lambda_3,\lambda_4$ in Equation 
(\ref{5_equation}) are empirically set as $1 \times 10^{-1}$. 
Besides, $\alpha$ and $\beta$ are chosen empirically as $1 \times 10^{-4}$ and 
$1 \times 10^{-1}$ respectively. 

\subsection{Evaluation metrics}

\begin{figure}[htb]
	\centering
	\includegraphics[width=0.45\textwidth]{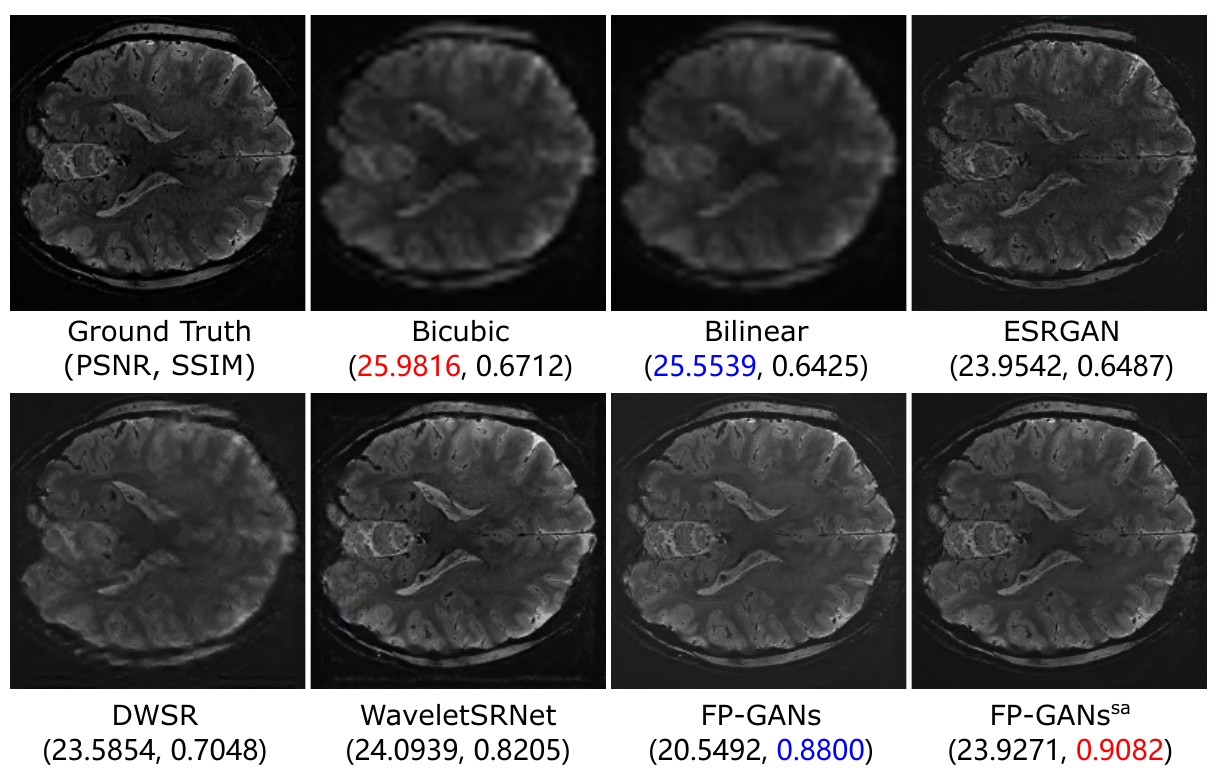}
	\caption{Comparison of quantitative value and visual quality. 
	Red and blue indicate the best and the second best performance, respectively.}
	\label{detail_psnr_experiment}
\end{figure}

Peak signal noise ratio (PSNR) and structural similarity index (SSIM) have been commonly 
applied in many super-resolution researches as standard evaluation metrics. 
Nevertheless, it was found that high PSNR doesn't guarantee a high visual quality during 
experiment. 
As illustrated in Figure \ref{detail_psnr_experiment}, Bicubic and Bilinear rank higher in PSNR, 
but the result of other methods are visually better from human perspective. 
Study\cite{7_ledig2017photo} also claimed that PSNR fails to assess image quality accurately 
with respect to human visual system. 
Under this circumstance, Fr\'echet Inception Distance (FID), calculating the feature distance 
of two images, is utilized to assess the perceptual quality of SR MR images. 
In summary, PSNR and SSIM are applied to measure the objective quality of images. 
FID is utilized to assess the visual quality from human perception.

\subsection{Performance}

Comprehensive analyses are carried out on MultiRes\_7T dataset with super-resolution 
scale factor $\times 4$. 
Figure \ref{result_scatter} reports the performance of FP-GANs and other competing methods 
upon SSIM and FID. 
It can be observed that FP-GANs achieves better both on SSIM and FID value than 
interpolation based methods (Bicubic and Bilinear), GAN based method 
(ESRGAN \cite{18_wang2018esrgan}) and wavelet based methods (DWSR \cite{13_guo2017deep} and 
WaveletSRNet \cite{12_huang2017wavelet}). 
Typically, equipped with sub-band attention, FP-GANs$^{sa}$ behaves a little better than 
FP-GANs. 

As illustrated in Figure \ref{single_detail}, the interpolation based methods produce 
over-smooth results. GAN based method---ESRGAN reproduces rich textures though, 
the result loses its coherency with the input. 
FP-GANs recovers finer anatomical structure with shaper and clearer textures 
while remaining consistency. 
To further examine the detail capture ability, difference heatmap, reflecting absolute 
difference between generated image and corresponding ground truth, is utilized 
as Figure \ref{Error_map} shows. It can be apparently observed that there is less 
difference in the heatmap of FP-GANs and FP-GANs$^{sa}$ than the competing methods. 
This confirms the detail sensitivity of the proposed structure. 

Besides, comparison with wavelet based methods (i.e. DWSR, WaveletSRNet) are 
conducted to investigate sub-band image reconstruction performance. As shown in 
Figure \ref{sunband_error_map}, sub-band generative adversarial networks (sub-band GANs) of 
FP-GANs reproduce more authentic sub-band images (LL, LH, HL, HH) in wavelet domain, 
and thus contributes to produce a compelling super-resolution image.

Table \ref{tab1} shows the overview of experiment quantitatively, from which it can be seen 
that FP-GANs outperforms the competing methods in most case, especially in $\times 4$ case.
Moreover, it's obvious that sub-band attention module exploited in FP-GANs$^{sa}$ enforces 
the model handle large scale super-resolution task better. From Figure \ref{wavelet_loss}, 
it can be concluded that wavelet loss contributes to stabilize and accelerate the convergence 
of FP-GANs.

\section{Ablation Study}
From Figure \ref{in_sa}, it can be seen that FP-GANs with instance normalization 
and sub-band attention mechanism converges most quickly and gets the highest score in SSIM. 
On the contrary, FP-GANs without the two module performs worst in SSIM. 
The ablation study on wavelet loss can be seen in Figure \ref{wavelet_loss}. 
It illustrates that wavelet loss contributes to stabilize and accelerate the convergence 
of FP-GANs. 

Form Figure \ref{sa_weight}, it can be observed that the weight parameters of 
sub-band attention gradually update from the initial random distribution to 
a regular distribution, 
where more attention are allocated on the high-frequency sub-bands, especially the HH one. 
This contributes to enhance the high-frequency anatomical textures of MR images.

\section{Discussion and Conclusion}

Losing textures is a universal problem of existing super-resolution methods. 
In this study, a novel super-resolution model---FP-GANs, integrated with wavelet 
transformation, is proposed to capture anatomical textures that are mainly in high-frequency. 
The proposed model divides the MR images into low-frequency and high-frequency sub-bands, 
and then conquer the super-resolution of each sub-band with a sub-band GAN separately. 
By treating the low-frequency global topology and high-frequency texture equally during 
super-resolution stage, it alleviates the detail insensitive problem. 
On the other hand, it contributes to stabilize training by simplifying the 
super-resolution task compared to the single GAN method like ESRGAN\cite{18_wang2018esrgan}. 
Besides, sub-band attention contributes to handle large scale super-resolution. 
Compared with interpolation and prior deep learning methods, FP-GANs shows its advantage 
on detail reconstruction, which is most important in medical image field. 



%

\appendices



\ifCLASSOPTIONcaptionsoff
  \newpage
\fi



\bibliographystyle{IEEEtran}
\bibliography{./FPGAN}
\end{document}